\RequirePackage{fix-cm}
\documentclass[11pt,a4paper]{article}

\usepackage[english]{babel}
\usepackage{amsmath,amssymb,amsthm,mathtools,physics}
\usepackage{bm}
\usepackage[numbers,sort&compress]{natbib}
\usepackage{graphicx}
\usepackage{xcolor}
\usepackage[normalem]{ulem}

\usepackage[a4paper,margin=25mm]{geometry}

\usepackage{authblk}

\usepackage{hyperref}

\usepackage{orcidlink}

\newcommand{\orcid}[1]{\orcidlink{#1}} 
\newtheorem{theorem}{Theorem}
\newtheorem{lemma}[theorem]{Lemma}

\newtheorem{proposition}[theorem]{Proposition}
\newtheorem{remark}{Remark}

\newcommand{\I}{\mathbb{I}}

\newcommand{\Cov}{\operatorname{Cov}}

\newcommand{\sgn}{\operatorname{sgn}}

\begin{document}

\title{Separating State Dependence from
Noncommutativity in Quantum Uncertainty Relations}

\author[1]{Haruki Yamashita\orcid{0009-0002-2439-0317}%
\thanks{Email: \texttt{haruki.yamashita1367@gmail.com}}}

\author[2]{Kenta Koshihara\orcid{0000-0002-3214-9315}%
\thanks{Email: \texttt{koshihara@aoni.waseda.jp}}}

\author[1]{Aina Mayumi\orcid{0009-0007-0956-1620}}

\author[3]{Gen Kimura\orcid{0000-0003-4288-2024}%
\thanks{Email: \texttt{gen.kimura.quant@gmail.com}}}

\affil[1]{Graduate School of Engineering and Science,
Shibaura Institute of Technology, Saitama, Japan}

\affil[2]{Department of Physics,
Waseda University, Tokyo, Japan}

\affil[3]{Graduate School of Information Sciences,
Tohoku University, Sendai, Japan}
\date{}
\maketitle
\vspace{-3.0em}

\begin{abstract}
The Robertson uncertainty relation establishes a fundamental connection between quantum uncertainty trade-offs and the noncommutativity of observables.
In this relation and many of its generalizations, however, noncommutativity enters through state-dependent quantities, and the resulting lower bounds can vanish for certain states even when the observables do not commute.
In this paper, we seek to separate state dependence from noncommutativity by constructing lower bounds that factorize into a coefficient depending only on the state and a measure of noncommutativity.
Using the squared Schatten-$p$ norm of the commutator as this measure, we derive a family of strengthened Robertson--Schr\"odinger uncertainty relations for finite-dimensional quantum systems in which the covariance and commutator terms are supplemented by a nonnegative term of this factorized form.
The coefficient of this additional term depends only on the two smallest eigenvalues of the density operator and is optimal for each fixed state, even when the Robertson--Schr\"odinger terms are retained.
For noncommuting observables in full-rank states, the additional term remains strictly positive even when the Robertson--Schr\"odinger lower bound vanishes, thereby quantifying ``hidden'' noncommutative uncertainty.
We further show that the trace-norm case ($p=1$) yields the strongest lower bound within the family, while all members reduce to the same exact identity for arbitrary qubit states and observables.

\end{abstract}

\noindent\textbf{Keywords:}
Quantum Physics, Uncertainty Relation, Schatten-$p$ Norm


\section{Introduction}
\label{section:Introduction}

The uncertainty principle is one of the fundamental features of quantum mechanics.
One of its most familiar quantitative expressions is the Heisenberg--Kennard uncertainty relation~\cite{Kennard1927},
\begin{equation}
V_\rho(x)V_\rho(p)\geq\frac{\hbar^2}{4},
\label{eq:HK}
\end{equation}
where $V_\rho(A)=\langle A^2\rangle_\rho-\langle A\rangle_\rho^2$ denotes the variance of an observable $A$ in a quantum state $\rho$, with $\langle A\rangle_\rho=\Tr(\rho A)$, and $\hbar$ is the reduced Planck constant.
Robertson subsequently generalized this relation to arbitrary pairs of observables $A$ and $B$~\cite{Robertson1929}:
\begin{equation}
V_\rho(A)V_\rho(B)
\geq\frac14|\langle[A,B]\rangle_\rho|^2.
\label{eq:Robertson}
\end{equation}
This relation explicitly connects the uncertainty trade-off between observables to their noncommutativity, characterized by the commutator $[A,B]=AB-BA$.
Since then, preparation uncertainty relations have been extensively generalized, including the Robertson--Schr\"odinger relation incorporating covariance~\cite{Schrodinger1930}, variance-sum relations~\cite{Huang2012,MacconePati2014}, and uncertainty relations based on quantum entropies~\cite{Deutsch1983,MaassenUffink1988,BertaChristandlColbeck2010}, skew information~\cite{Luo2005}, and Fisher information~\cite{Hall2001,ChiewGessner2022}.
See also Refs.~\cite{WehnerWinter2010,Englert2024} for reviews.

A distinctive feature of the Heisenberg--Kennard relation~\eqref{eq:HK} is that its lower bound is independent of the quantum state.
It guarantees that no quantum state can simultaneously eliminate the uncertainties of position and momentum.
In contrast, the lower bound of the Robertson relation generally depends on the quantum state and can vanish even for noncommuting observables.
This limitation persists in the Schr\"odinger relation, 

\begin{equation}
V_\rho(A)V_\rho(B)
\geq
\frac14|\langle [A,B]\rangle_\rho|^2
+
\mathrm{Cov}_\rho(A,B)^2, 
\label{eq:RobertsonSchrodinger}
\end{equation}
where $\mathrm{Cov}_\rho(A,B)=\frac12\langle AB+BA\rangle_\rho-\langle A\rangle_\rho\langle B\rangle_\rho$ denotes the symmetric quantum covariance. 
Its lower bound can vanish even for a full-rank mixed state despite a strictly positive variance product.
Several direct refinements have been developed so far~\cite{Nagaoka2005,Hayashi2017,Park2005,ZhengZhang2017,KimuraMayumiOhno2026,KimuraMayumiYamashita2026}, and 
in particular, Ref.~\cite{KimuraMayumiOhno2026} established the following strengthened Robertson--Schr\"odinger uncertainty relation:
\begin{equation}
V_\rho(A)V_\rho(B)
\geq
\frac14|\langle[A,B]\rangle_\rho|^2
+
\mathrm{Cov}_\rho(A,B)^2
+
\frac{\lambda_1(\rho)\lambda_2(\rho)}
{\lambda_1(\rho)+\lambda_2(\rho)}
\langle|[A,B]|^2\rangle_\rho,
\label{eq:KimuraMayumiOhno}
\end{equation}
where $\lambda_1(\rho)$ and $\lambda_2(\rho)$ are the two smallest eigenvalues of $\rho$.
For full-rank states, the additional term is strictly positive for any pair of noncommuting observables, revealing a noncommutative trade-off even when the Robertson--Schr\"odinger lower bound vanishes.
However, the factor $\langle|[A,B]|^2\rangle_\rho$ itself also depends on both the observables and the quantum state, so the state dependence remains intertwined with the measure of noncommutativity.

This naturally raises the question of whether one can establish a Kennard-like, state-independent positive lower bound for arbitrary noncommuting observables.
Such a bound, however, cannot hold in finite-dimensional systems: an eigenstate of one observable has zero variance, while the variance of the other remains finite.
The same obstruction applies to bounded observables in infinite-dimensional systems by considering approximate eigenstates.
Thus, a strictly positive, state-independent lower bound valid for all states, such as the Heisenberg--Kennard bound, is possible only for a pair of unbounded observables.

Motivated by this observation, we take a different approach.
Rather than eliminating state dependence, we seek to separate it from the noncommutativity of observables.
More precisely, we consider uncertainty relations of the form
\begin{equation}
V_\rho(A)V_\rho(B)
\geq c(\rho)\|[A,B]\|^2,
\label{eq:ObjUR1}
\end{equation}
where $\|\cdot\|$ is a suitable norm that quantifies the noncommutativity of the observables.
Importantly, the coefficient $c(\rho)$ depends only on the quantum state, whereas $\|[A,B]\|$ quantifies solely the degree of noncommutativity of the observables, independently of the state.
Our central problem is to determine, for each fixed state $\rho$ and a given norm, the optimal coefficient $c^{\mathrm{opt}}(\rho)$, namely, the largest coefficient for which the inequality holds for all pairs of observables $A$ and $B$.

In this work, we address this problem for the family of Schatten-$p$ norms $\|\cdot\|_p$ with $1\leq p\leq\infty$, which includes the trace norm ($p=1$), the Hilbert--Schmidt norm ($p=2$), and the operator norm ($p=\infty$) and solve this optimization problem for every Schatten-$p$ norm.

Specifically, for every $1\leq p\leq\infty$, we show that the optimal coefficient for each fixed state $\rho$ is determined solely by its two smallest eigenvalues $\lambda_1(\rho)$ and $\lambda_2(\rho)$, obtaining the sharp uncertainty relation
\begin{equation}\label{eq:MainUR1}
V_\rho(A)V_\rho(B)
\geq
\frac{\lambda_1(\rho)\lambda_2(\rho)}{4^{1/p}}
\|[A,B]\|_p^2.
\end{equation}
Surprisingly, our result provides more than an alternative Robertson-type bound: the same commutator contribution can be added to the full Robertson--Schr\"odinger lower bound without reducing the optimal coefficient, yielding
\begin{equation}\label{eq:MainUR}
V_\rho(A)V_\rho(B)
\geq \underbrace{\frac14|\langle[A,B]\rangle_\rho|^2}_{\text{Robertson term}}+\underbrace{\mathrm{Cov}_\rho(A,B)^2}_{\text{Schr\"odinger term}} +
\underbrace{\frac{\lambda_1(\rho)\lambda_2(\rho)}{4^{1/p}}
\|[A,B]\|_p^2}_{\text{New Trade-off}},
\end{equation}
The additional third term is strictly positive for any pair of noncommuting observables for any full-rank state.
It thus quantifies a ``hidden'' noncommutative contribution to quantum uncertainty that is not captured by the conventional Robertson--Schr\"odinger lower bound. 
Moreover, 
the coefficient remains optimal even in the presence of the Robertson--Schr\"odinger terms.
We further prove that the lower bound in the family~\eqref{eq:MainUR} is maximized at $p=1$ over the entire range $1\leq p\leq\infty$.
The optimal uncertainty relation within this family is therefore
\begin{equation}\label{eq:URTR}
V_\rho(A)V_\rho(B)
\geq
\frac14|\langle[A,B]\rangle_\rho|^2
+\mathrm{Cov}_\rho(A,B)^2
+\frac{\lambda_1(\rho)\lambda_2(\rho)}{4}
\bigl(\Tr|[A,B]|\bigr)^2.
\end{equation}

Finally, and remarkably, for qubit systems the additional term becomes independent of $p$, and all members of our family of uncertainty relations~\eqref{eq:MainUR} reduce to the same exact identity for arbitrary states and observables.

The present results parallel the structure of the uncertainty relations \eqref{eq:KimuraMayumiOhno}. 
That work introduced an additional trade-off term proportional to the expectation value of the squared modulus of the commutator.
It showed that this term provides a sharp lower bound on the variance product and can also be added to the Robertson--Schr\"odinger lower bound without changing its optimal coefficient.
For qubit systems, the resulting strengthened relation becomes an exact identity for arbitrary states and observables.
In Eq.~\eqref{eq:KimuraMayumiOhno}, the commutator factor $\langle|[A,B]|^2\rangle_\rho$ generally depends on both the observables and the state.
By contrast, the additional term in our relations \eqref{eq:MainUR} \emph{separates these two dependencies}: all state dependence is confined to the coefficient $\lambda_1(\rho)\lambda_2(\rho)/4^{1/p}$, while the squared Schatten-$p$ norm of the commutator $\|[A,B]\|_p^2$ quantifies noncommutativity independently of the state.

This paper is organized as follows. 
In Sec.~\ref{section:ProofMainTheorem}, we present the new generalization of the uncertainty relation including state-independent Schatten-$p$ norm of commutator, together with a rigorous proof of it, and show that the trace-norm case $p=1$ provides the strongest bound within the Schatten norm family.
We analytically and numerically compare our result in Sec.~\ref{section:TightnessOfGeneralizedUncertaintyRelations} with the above two generalized uncertainty relations, in particular showing that our result also becomes the exact equality in qubit systems. 
We conclude this paper in Sec.~\ref{section:Conclusions}. 
Several technical calculations concerning the proof of the main result are presented in Appendices~\ref{appendix:CalculatingWeightedNorm}-\ref{appendix:ProofOfTraceNormDominanceGeneralized}.

Throughout this paper, we consider a quantum system described by a finite-dimensional Hilbert space $\mathcal H\simeq\mathbb C^d$.
An observable is represented by a Hermitian operator $A=A^\dagger$ on $\mathcal H$.
A quantum state is represented by a density operator $\rho$, namely a positive semidefinite operator $\rho\geq0$ with unit trace $\Tr\rho=1$.
The sets of all linear operators, Hermitian operators, and density operators on $\mathcal H$ are denoted by $\mathcal L(\mathcal H)$, $\mathcal L_h(\mathcal H)$, and $\mathcal S(\mathcal H)$, respectively. 
For a Hermitian operator $A\in\mathcal L_h(\mathcal H)$, we denote its eigenvalues, counted with multiplicity and arranged in nondecreasing order, by 
\[
\lambda_1(A)\leq\lambda_2(A)\leq\cdots\leq\lambda_d(A). 
\]


\section{Optimal uncertainty relations separating state dependence from noncommutativity}
\label{section:ProofMainTheorem}

In this section, we formulate and prove the uncertainty relations \eqref{eq:MainUR1}, which separates state dependence from observable noncommutativity through an optimal coefficient and a norm of the commutator. 
We begin by recalling the Schatten-$p$ norms.
For a linear operator $X\in\mathcal L(\mathcal H) \simeq M_d(\mathbb C)$, the operator Schatten norm is defined by~\cite{Bhatia1997,Simon2005}

\begin{equation}
\|X\|_p
=\bm(\Tr|X|^p\bm)^{1/p}, \quad 1\leq p \leq \infty,
\label{eq:SchattenNorm}
\end{equation}
where $|X| = \sqrt{X^\dagger X}$. 
Using an eigenvalue decomposition of $|X|$, $|X|=\sum_{i=1}^d\sigma_i(X)|\psi_i\rangle\langle\psi_i|$, where $\sigma_i(X)\geq0$ are the singular values of $X$, we obtain
$\|X\|_p=\left(\sum_{i=1}^d\sigma_i(X)^p\right)^{1/p}$ for $1\leq p<\infty$.
Thus, the Schatten-$p$ norm of $X$ is the $\ell_p$ norm of its vector of singular values $(\sigma_1(X),\ldots,\sigma_d(X))$.
In particular, the Schatten-$p$ norm~\eqref{eq:SchattenNorm} reduces to the trace norm $\|X\|_1=\Tr|X|$ for $p=1$ and the Hilbert--Schmidt norm $\|X\|_2=\sqrt{\Tr(X^\dagger X)}$ for $p=2$.
In the limit $p\to\infty$, it becomes the operator (or spectral) norm
$\|X\|_\infty=\max_{1\leq i\leq d}\sigma_i(X)=\max_{\|\psi\|=1}\|X\psi\|$,
where $\|\psi\|=\sqrt{\langle\psi|\psi\rangle}$ denotes the Hilbert space norm.

The main results of this paper are a family of sharp uncertainty relations formulated in terms of the Schatten-$p$ norms.
We first establish the following optimal bound of the form~\eqref{eq:ObjUR1}, which separates state dependence from observable noncommutativity.
\begin{proposition}
\label{prop:MainTheorem}
For each $p\in[1,\infty]$, the following uncertainty relation holds
\begin{equation}
V_\rho(A)V_\rho(B)
\geq
\frac{\lambda_1(\rho)\lambda_2(\rho)}{4^{1/p}}
\|[A,B]\|_p^2,
\label{eq:FactorizedUR}
\end{equation}
for any quantum state $\rho\in\mathcal S(\mathcal H)$ and any pair of observables $A,B\in\mathcal L_h(\mathcal H)$.
Here, $\lambda_1(\rho)$ and $\lambda_2(\rho)$ are the two smallest eigenvalues of $\rho$, counted with multiplicity.
For each fixed state $\rho$ and each $p\in[1,\infty]$, the coefficient $\lambda_1(\rho)\lambda_2(\rho)/4^{1/p}$ is optimal in the sense that it is the largest coefficient for which the inequality holds for all pairs of observables $A$ and $B$.
\end{proposition}
We further show that the commutator contribution in Eq.~\eqref{eq:FactorizedUR} can be added to the Robertson--Schr\"odinger lower bound without reducing its optimal coefficient, yielding the following strengthened uncertainty relation.

\begin{theorem}
\label{theorem:MainTheorem}
For each $p\in[1,\infty]$, the following uncertainty relation holds
\begin{equation}
V_\rho(A)V_\rho(B)
\geq
\frac14|\langle[A,B]\rangle_\rho|^2
+
\mathrm{Cov}_\rho(A,B)^2
+
\frac{\lambda_1(\rho)\lambda_2(\rho)}{4^{1/p}}
\|[A,B]\|_p^2,
\label{eq:MainResult}
\end{equation}
for any quantum state $\rho\in\mathcal S(\mathcal H)$ and any pair of observables $A,B\in\mathcal L_h(\mathcal H)$.
The coefficient $\lambda_1(\rho)\lambda_2(\rho)/4^{1/p}$ remains optimal for each fixed state $\rho$ and each $p\in[1,\infty]$.
\end{theorem}
In the following, we first prove the inequalities in Proposition~\ref{prop:MainTheorem} and Theorem~\ref{theorem:MainTheorem} for $p=1$, and then extend them to all $p\in[1,\infty]$.
In the process of this extension, we also observe that the relation~\eqref{eq:URTR} for the trace-norm case $p=1$ provides the strongest lower bound within each family.
We subsequently establish the optimality of the coefficients for each fixed state and each $p$ by constructing noncommuting observables that attain equality.

The following two lemmas provide the key estimates for the case $p=1$.
The first follows from Cauchy's interlacing theorem and accounts for the product $\lambda_1(\rho)\lambda_2(\rho)$ appearing in the optimal coefficient.

\begin{lemma}
\label{theorem:CalculatingWeightedNorm}
Let $\omega>0$ be a positive-definite operator and let $\lambda_1(\omega)$ and $\lambda_2(\omega)$ denote its two smallest eigenvalues, counted with multiplicity. 
For any linear operator $X\in\mathcal L(\mathcal H)$ and an orthonormal basis $\{|\phi_1\rangle,|\phi_2\rangle,\ldots,|\phi_d\rangle\}$ on $\mathcal H$, we have
\begin{equation}
\langle |X|^2\rangle_\omega
\geq
\lambda_1(\omega)\lambda_2(\omega)
\sum_{i\neq j}
\frac{|X_{ij}|^2}{\omega_{ii}}
.
\end{equation}
where $X_{ij}=\langle\phi_i|X|\phi_j\rangle$
and $\omega_{ii}=\langle\phi_i|\omega|\phi_i\rangle>0$
are the matrix elements in the basis.
\end{lemma}
The proof is given in Appendix~\ref{appendix:CalculatingWeightedNorm}. 
Using this lemma, we obtain the following:  
\begin{lemma}
\label{theorem:TraceNormBound}
Let $\omega>0$ be a positive-definite operator, and let $X\in\mathcal L(\mathcal H)$ and $Y\in\mathcal L_h(\mathcal H)$ satisfy
\begin{equation}
[X,Y]=[X^\dagger,Y].
\label{eq:HermitianCommutatorCondition}
\end{equation}
Then
\begin{equation}
\langle |X|^2\rangle_\omega
\langle|Y|^2\rangle_\omega
\geq
\frac{\lambda_1(\omega)\lambda_2(\omega)}{4}
\|[X,Y]\|_1^2
.
\end{equation}
\end{lemma}

\begin{proof}

The assumption~(\ref{eq:HermitianCommutatorCondition}) implies that
\begin{equation}
H=\mathrm i[X,Y]
\end{equation}
is a Hermitian operator. 
Its polar decomposition can be therefore obtained with a Hermitian unitary operator: 
\begin{equation}
H=U|H|,
\quad
U^\dagger U=\I
\quad
\text{and}
\quad
U^\dagger=U.
\end{equation}
To construct $U$, consider an eigenvalue decomposition $H=\sum_{i=1}^d\lambda_i(H)|\psi_i\rangle\langle \psi_i|$, which gives  $|H|=\sum_{i=1}^d|\lambda_i(H)||\psi_i\rangle\langle \psi_i|$. 
We then define $U=\sum_{i=1}^d\sgn\bigl(\lambda_i(H)\bigr)|\psi_i\rangle\langle\psi_i|$ where $\sgn(x)=1$ for $x\geq0$ and $\sgn(x)=-1$ for $x<0$.
Since every eigenvalue of $U$ is either $1$ or $-1$, $U$ is both Hermitian and unitary, and it satisfies $H=U|H|$.
In particular, the convention $\sgn(0)=1$ ensures that $U$ remains unitary even when $H$ has zero eigenvalues.

Moreover, the assumption~(\ref{eq:HermitianCommutatorCondition}) constrains the matrix elements of $X$ expanded with eigenvectors of $Y$; using an eigenvalue decomposition 
\begin{equation}
Y
=\sum_{i=1}^d
\lambda_i(Y)|\phi_i\rangle\langle \phi_i|
\end{equation}
and denoting $X_{ij}=\langle\phi_i|X|\phi_j\rangle$ as the matrix element of $X$,  the assumption~(\ref{eq:HermitianCommutatorCondition}) leads to
\begin{equation}
0=
\langle\phi_i|[X-X^\dagger,Y]|\phi_j\rangle
=
(X_{ij}-X_{ji}^\ast)
\bm(\lambda_j(Y)-\lambda_i(Y)\bm)
,
\end{equation}
which implies
\begin{equation}
\lambda_i(Y)\neq \lambda_j(Y)
\quad
\Rightarrow
\quad
X_{ij}=X_{ji}^\ast
.
\label{eq:ConditionalHemitianity}
\end{equation}
Denoting the matrix elements $H_{ij}=\langle\phi_i|H|\phi_j\rangle$,  $U_{ij}=\langle\phi_i|U|\phi_j\rangle$, and $\omega_{ii}=\langle\phi_i|\omega|\phi_i\rangle>0$, we can evaluate the trace norm $\|[X,Y]\|_1$ as
\begin{align}
\|[X,Y]\|_1
=
|\Tr(|H|)|
=
|\Tr(UH)|
\leq{}&
\sum_{i,j=1}^d
|U_{ij}||H_{ji}|=
\sum_{i,j=1}^d
|U_{ij}||\lambda_i(Y)-\lambda_j(Y)|\,|X_{ji}|.
\nonumber\\
={}&
\sum_{\lambda_i(Y)\neq\lambda_j(Y)}
|U_{ij}||X_{ij}||\lambda_i(Y)-\lambda_j(Y)|,
\end{align}
where the last equality uses~\eqref{eq:ConditionalHemitianity}.
Here and below, the sum runs over all ordered pairs $(i,j)$ with $1\leq i,j\leq d$ such that $\lambda_i(Y)\neq\lambda_j(Y)$.
Applying the Cauchy--Schwarz inequality with positive weights $\omega_{ii}^{-1}+\omega_{jj}^{-1}$, we obtain
\begin{align}
{}&
\sum_{\lambda_i(Y)\neq\lambda_j(Y)}
|U_{ij}||X_{ij}|
|\lambda_i(Y)-\lambda_j(Y)|
\nonumber\\
{}&\quad\leq
\sqrt{
\sum_{\lambda_i(Y)\neq\lambda_j(Y)}
\biggl(
\frac{1}{\omega_{ii}}+\frac{1}{\omega_{jj}}
\biggr)
|X_{ij}|^2
}
\sqrt{
\sum_{\lambda_i(Y)\neq\lambda_j(Y)}
\biggl(
\frac{1}{\omega_{ii}}+\frac{1}{\omega_{jj}}
\biggr)^{-1}
|U_{ij}|^2|\lambda_i(Y)-\lambda_j(Y)|^2
}
.
\label{eq:WeightedCauchySchwarz}
\end{align}
Using the conditional Hermiticity~(\ref{eq:ConditionalHemitianity}) of $X$, together with Lemma~\ref{theorem:CalculatingWeightedNorm}, 
we bound the first factor on the right-hand side of~\eqref{eq:WeightedCauchySchwarz} as follows:
\begin{equation}
\sqrt{
\sum_{\lambda_i(Y)\neq\lambda_j(Y)}
\biggl(
\frac{1}{\omega_{ii}}+\frac{1}{\omega_{jj}}
\biggr)
|X_{ij}|^2
}
=
\sqrt{
\sum_{\lambda_i(Y)\neq\lambda_j(Y)}
\frac{2|X_{ij}|^2}{\omega_{ii}}
}
\leq
\sqrt{
\sum_{i\neq j}
\frac{2|X_{ij}|^2}{\omega_{ii}}
}
\leq
\sqrt{\frac{2\langle |X|^2\rangle_\omega}{\lambda_1(\omega)\lambda_2(\omega)}}
.
\label{eq:XNormEvaluation}
\end{equation}
A straightforward calculation, given in Appendix~\ref{appendix:ProofOfYNormEvaluation}, yields the following bound on the second factor:
\begin{equation}
\sqrt{
\sum_{\lambda_i(Y)\neq\lambda_j(Y)}
\biggl(
\frac{1}{\omega_{ii}}+\frac{1}{\omega_{jj}}
\biggr)^{-1}
|U_{ij}|^2|\lambda_i(Y)-\lambda_j(Y)|^2
}
\leq
\sqrt{2\langle |Y|^2\rangle_\omega}
.
\label{eq:YNormEvaluation}
\end{equation}
Combining the above estimates, we obtain
$\|[X,Y]\|_1^2\leq \frac{4\langle|X|^2\rangle_\omega\langle|Y|^2\rangle_\omega}{\lambda_1(\omega)\lambda_2(\omega)}$,
which completes the proof.
\end{proof}
For faithful state $\rho>0$, the inequality in Proposition~\ref{prop:MainTheorem} for $p=1$ follows immediately by applying lemma~\ref{theorem:TraceNormBound} with $\omega=\rho$, $X=A-\langle A\rangle_\rho\I$, and $Y=B-\langle B\rangle_\rho\I$, noting that $[X,Y]=[A,B]$.
For rank-deficient states, the inequality is trivial because $\lambda_1(\rho)=0$. 

We next prove the inequality in Theorem~\ref{theorem:MainTheorem} for $p=1$ using the method of Ref.~\cite{KimuraMayumiOhno2026}.
Let $\omega>0$ and $A,B\in\mathcal L_h(\mathcal H)$.
If $B=0$, the inequality below is trivial.
Assume $B\neq0$.
For any $z\in\mathbb C$, the operators $X=A+zB$ and $Y=B$ satisfy
\begin{equation}
[X,Y]=[X^\dagger,Y]=[A,B],
\end{equation}
so Lemma~\ref{theorem:TraceNormBound} applies.
Choosing
\begin{equation}
z=-\frac{\langle AB\rangle_\omega^\ast}{\langle|B|^2\rangle_\omega},
\end{equation}
which is well-defined since $\omega>0$ and $B\neq0$, gives
\begin{equation}
\langle|A+zB|^2\rangle_\omega
=
\langle|A|^2\rangle_\omega
-
\frac{|\langle AB\rangle_\omega|^2}{\langle|B|^2\rangle_\omega}.
\end{equation}
We therefore obtain
\begin{equation}
\langle|A|^2\rangle_\omega
\langle|B|^2\rangle_\omega
-
|\langle AB\rangle_\omega|^2
\geq
\frac{\lambda_1(\omega)\lambda_2(\omega)}{4}
\|[A,B]\|_1^2.
\label{eq:TraceNormBoundCorollary}
\end{equation}

Now let $\rho>0$ and introduce the shifted observables
\begin{equation}
\hat A=A-\langle A\rangle_\rho\I,
\quad
\hat B=B-\langle B\rangle_\rho\I.
\end{equation}
Then $\langle|\hat A|^2\rangle_\rho=V_\rho(A)$, $\langle|\hat B|^2\rangle_\rho=V_\rho(B)$, and $[\hat A,\hat B]=[A,B]$.
Moreover, using $AB=\frac12[A,B]+\frac12\{A,B\}$, we obtain
\begin{align}
|\langle\hat A\hat B\rangle_\rho|^2
&=
\left|
\mathrm{Cov}_\rho(A,B)
+\frac12\langle[A,B]\rangle_\rho
\right|^2
\nonumber\\
&=
\mathrm{Cov}_\rho(A,B)^2
+\frac14|\langle[A,B]\rangle_\rho|^2,
\end{align}
because $\mathrm{Cov}_\rho(A,B)$ is real and $\langle[A,B]\rangle_\rho$ is purely imaginary.
Applying~\eqref{eq:TraceNormBoundCorollary} to $\hat A$ and $\hat B$ with $\omega=\rho$ proves the inequality in Theorem~\ref{theorem:MainTheorem} for $p=1$.
For rank-deficient states, the additional term vanishes, and the inequality reduces to the Robertson--Schr\"odinger relation.

Next, the inequalities for $p\in(1,\infty]$ follow from the case $p=1$ via the following lemma, which establishes the dominance of the trace norm over appropriately rescaled Schatten-$p$ norms.
\begin{lemma}
\label{theorem:TraceNormDominance}
Let $H\in\mathcal L_h(\mathcal H)$ be a traceless Hermitian operator, i.e.,  $\Tr H=0$ and $H^\dagger=H$, 
then
\begin{equation}
\|H\|_1
\geq
2^{1-1/p}\|H\|_p
,
\quad
p\in[1,\infty]
.
\label{eq:TraceNormDominance}
\end{equation}
\end{lemma}

Before proving the lemma, we distinguish~\eqref{eq:TraceNormDominance} from the standard monotonicity of Schatten norms~\cite{Bhatia1997},
$\|X\|_1\geq\|X\|_p\geq\|X\|_{p'}\geq\|X\|_\infty$ for $1\leq p\leq p'\leq\infty$,
which holds for any linear operator $X\in\mathcal L(\mathcal H)$.
The factor $2^{1-1/p}$ in~\eqref{eq:TraceNormDominance} increases with $p$, whereas $\|H\|_p$ decreases.
Consequently, the dominance of the trace norm over these rescaled Schatten norms does not follow from standard monotonicity alone.
It relies on the balance between the positive and negative spectral parts imposed by the tracelessness of $H$.
For general linear operators, a related inequality shows that the operator norm can instead dominate after rescaling.
We discuss this extension in Appendix~\ref{appendix:ProofOfTraceNormDominanceGeneralized}, together with the special structure of traceless Hermitian operators in qubit and qutrit systems.

\begin{proof}
Note that the traceless property excludes both $H\geq0$ and $H\leq0$ cases. 
Let us consider a nontrivial case $H\neq0$, by applying an eigenvalue decomposition 
$H
=
\sum_{i=1}^d
\lambda_i(H)
|\psi_i\rangle\langle\psi_i|
$
with
$\lambda_1(H)
\leq
\cdots
\leq
\lambda_k(H)
<0
<
\lambda_{k+1}(H)
\leq
\cdots
\leq
\lambda_d(H)$.
Then we can divide $H=H_+-H_-$, whose positive part $H_+$ and negative one $H_-$ are given by
\begin{equation}
H_+=\sum_{i=k+1}^d
\lambda_i(H)|\psi_i\rangle\langle \psi_i|
>0
,
\quad
H_-=\sum_{i=1}^k
|\lambda_i(H)||\psi_i\rangle\langle \psi_i|
>0
.
\end{equation}
So the traceless property $\Tr H=0$, together with $|H|=H_++H_-$, implies
\begin{equation}
\Tr H_+=\Tr H_-=\frac12\|H\|_1.
\label{eq:TracelessHermitianSymmetry}
\end{equation}

Then the Schatten norm within $1<p<\infty$ can be evaluated as
\begin{align}
\|H\|_p^p
={}&
\sum_{i=k+1}^d
\lambda_i(H)^p
+
\sum_{i=1}^k
|\lambda_i(H)|^p
\nonumber\\
={}&
\sum_{i=k+1}^d
\lambda_i(H)
\lambda_i(H)^{p-1}
+
\sum_{i=1}^k
|\lambda_i(H)|
|\lambda_i(H)|^{p-1}
\nonumber\\
\leq{}&
\sum_{i=k+1}^d
\lambda_i(H)
\biggl(
\sum_{j=k+1}^d
\lambda_j(H)
\biggr)^{p-1}
+
\sum_{i=1}^k
|\lambda_i(H)|
\biggl(\sum_{j=1}^k
|\lambda_j(H)|
\biggr)^{p-1}
\nonumber\\
={}&
(\Tr H_+)^{p}
+
(\Tr H_-)^{p}
=
\frac{2}{2^p}\|H\|_1^p
,
\end{align}
where the inequality follows from $\sum_i a_i^p\leq(\sum_i a_i)^p$ for nonnegative $a_i$ and $p>1$, applied separately to the positive and negative spectral parts.
Taking the $p$th root gives the desired inequality $2^{1-1/p}\|H\|_p\leq\|H\|_1$. 
The case $p=\infty$ follows by taking the limit $p\to\infty$, since $\|H\|_p\to\|H\|_\infty$ and $2^{1-1/p}\to2$.
\end{proof}

Applying Lemma~\ref{theorem:TraceNormDominance} to the traceless Hermitian operator $H=i[A,B]$, we obtain Eq.~\eqref{eq:MainResult} for all $p\in[1,\infty]$ from the case of $p=1$ established above.
This argument also shows that $p=1$ provides the strongest lower bound within this family.
Thus, the inequality in Theorem~\ref{theorem:MainTheorem} has been established for all $p\in [1,\infty]$.
It remains to show that the coefficient is optimal.

We next show the optimality of the coefficient in Theorem~\ref{theorem:MainTheorem} for any fixed full-rank state $\rho$.
Let $\ket{1}$ and $\ket{2}$ be orthonormal eigenvectors of $\rho$ corresponding to its smallest eigenvalue $\lambda_1(\rho)$ and $\lambda_2(\rho)$, respectively.
We construct the following two observables:
\begin{align}
    A^{\ast} 
    &:= 
    \lambda_2(\rho)\ket{1}\bra{1}-\lambda_1(\rho)\ket{2}\bra{2},\\
    B^{\ast}
    &:=
    \ket{1}\bra{2}+\ket{2}\bra{1}.
\end{align}
Since $\rho$ is diagonal in the eigenbasis, we have $\langle A^{\ast}\rangle_{\rho} =\langle B^{\ast}\rangle_{\rho}=0$,
and hence
\begin{align}
    V_{\rho}(A^{\ast})
    &=
\lambda_1(\rho)\lambda_2(\rho)\bigl(\lambda_1(\rho)+\lambda_2(\rho)\bigr),\quad
    V_{\rho}(B^{\ast})
    =
    \lambda_1(\rho)+\lambda_2(\rho),\\
    &[A^{\ast},B^{\ast}]
    =
\bigl(\lambda_1(\rho)+\lambda_2(\rho)\bigr)\bigl(\ket{1}\bra{2}-\ket{2}\bra{1}\bigr).
\end{align}
    Thus, $[A^{\ast},B^{\ast}]$ has exactly two nonzero singular values, both equal to $\lambda_1(\rho)+\lambda_2(\rho)$.
    Consequently, for every $p\in[1,\infty]$, 
\begin{equation}
    \|[A^{\ast},B^{\ast}]\|_p^2
    =
4^{1/p}\bigl(\lambda_1(\rho)+\lambda_2(\rho)\bigr)^2.
\end{equation}
Moreover, both $[A^{\ast},B^{\ast}]$ and $\{A^{\ast},B^{\ast}\}$ are off-diagonal in the eigenbasis of $\rho$, which implies
\begin{equation}
    \langle [A^{\ast},B^{\ast}]\rangle_{\rho}=\mathrm{Cov}_{\rho}(A^{\ast},B^{\ast})=0
\end{equation}
Combining these equalities, we obtain
\begin{equation}
    V_{\rho}(A^{\ast})V_{\rho}(B^{\ast})=
    \frac{\lambda_1(\rho)\lambda_2(\rho)}{4^{1/p}}
\|[A^{\ast},B^{\ast}]\|_p^2.
\end{equation}
Thus, Eq.~\eqref{eq:MainResult} is saturated by this pair of observables for every $p\in[1,\infty]$.
In particular, equality is attained for a pair of observables with a nonzero commutator, while both terms in the Robertson--Schr\"odinger bound vanish. 
To show that the coefficient is optimal, let $c_p(\rho)$ be any coefficient that can replace $\lambda_1(\rho)\lambda_2(\rho)/4^{1/p}$ in Eq.~\eqref{eq:MainResult} while preserving its validity for all Hermitian observables $A$ and $B$. 
Substituting the above observables into this inequality, we obtain
\begin{equation}
    c_p(\rho)\le \frac{V_{\rho}(A^{\ast})V_{\rho}(B^{\ast})}{\|[A^{\ast},B^{\ast}]\|_p^2}=\frac{\lambda_1(\rho)\lambda_2(\rho)}{4^{1/p}}.
\end{equation}
Thus, this establishes the optimality of the coefficient for full-rank states.
For rank-deficient states, take a unit vector $\ket{1}$ in the kernel of $\rho$ and a unit vector $\ket{2}$ orthogonal to $\ket{1}$.
Setting $A^{\ast}=\ket{1}\bra{1}$ and $B^{\ast}=\ket{1}\bra{2}+\ket{2}\bra{1}$, we have $V_{\rho}(A^{\ast})=0$ and $[A^{\ast},B^{\ast}]\neq 0$. Therefore, the optimal coefficient is zero.
Hence, the coefficient $\frac{\lambda_1(\rho)\lambda_2(\rho)}{4^{1/p}}$ is optimal for every quantum state $\rho$.
This completes the proof of Theorem~\ref{theorem:MainTheorem}. The same construction can also show the optimality of the coefficient in Proposition~\ref{prop:MainTheorem}.


\section{Tightness of the optimal uncertainty relations}
\label{section:TightnessOfGeneralizedUncertaintyRelations}

In the previous section, we established the optimal uncertainty relations by separating the state dependence from the noncommutativity of the observables. 
In this section, we first show that our bound becomes an exact equality in qubit systems. We then compare it with that of existing uncertainty relations in higher-dimensional systems.


\subsection{Exact equalities in qubit systems}

Several refinements of the Robertson--Schr\"odinger uncertainty relation have been proposed.
Among them, the Nagaoka--Hayashi uncertainty relation~\cite{Nagaoka2005,Hayashi2017}  provides a
particularly strong lower bound:
\begin{equation}\label{eq:NagaokaHayashi}
    V_{\rho}(A)V_{\rho}(B)
    \ge
    \frac{1}{4}\|\sqrt{\rho}[A,B]\sqrt{\rho}\|_1^2+\Cov_{\rho}(A,B)^2.
\end{equation}
Since 
\begin{equation}
    |\langle[A,B]\rangle_{\rho}| \le \|\sqrt{\rho}[A,B]\sqrt{\rho}\|_1,
\end{equation}
this relation is always at least as strong as the Robertson--Schr\"{o}dinger uncertainty relation.
However, its commutator term is not a simple measurement statistics like an expectation. Therefore, it is difficult to evaluate experimentally.

Interestingly, in qubit systems ($d=2$), the Nagaoka--Hayashi relation~\eqref{eq:NagaokaHayashi}, the relation~\eqref{eq:KimuraMayumiOhno}, and our Schatten-$p$ uncertainty relations~\eqref{eq:MainResult} all become exact equalities for arbitrary states and observables.
The equality for the Nagaoka--Hayashi relation is known (see Ref.~\cite{Hayashi2017}), and that for the KMOLC relation was shown in Ref.~\cite{KimuraMayumiOhno2026}.

    The exactness of our bound follows from a special property of qubit commutators. 
    Since $i[A,B]$ is a traceless Hermitian operator in dimension two, its eigenvalues are $\sigma$ and $-\sigma$ for some $\sigma\geq 0$. 
    Thus, its two singular values are equal, and $|[A,B]|=\sigma I$. 
    Therefore, we have $\|[A,B]\|_p=2^{1/p}\sigma$ for every $p\in [1,\infty]$ and $\langle|[A,B]|^2\rangle_{\rho}=\sigma^2$.
    It follows that

\begin{equation}\label{eq:QubitSchattenEquality}
        \frac{\lambda_1(\rho)\lambda_2(\rho)}{4^{1/p}}\|[A,B]\|_p^2
        =
        \frac{\lambda_1(\rho)\lambda_2(\rho)}{4}\|[A,B]\|_1^2
        =
        \lambda_1(\rho)\lambda_2(\rho)\langle|[A,B]|^2\rangle_{\rho}
        \quad
        \forall p\in[1,\infty]
\end{equation}
    Since $\lambda_1(\rho)+\lambda_2(\rho)=1$, the last term is exactly the additional term in Eq.~\eqref{eq:KimuraMayumiOhno}.
    Therefore, our bound~\eqref{eq:MainResult} coincides with Eq.~\eqref{eq:KimuraMayumiOhno} and also becomes an equality for all qubit states and observables. 
By rewriting the product of eigenvalues of $\rho$ in its purity $P(\rho)=\Tr\rho^2=1-2\lambda_1(\rho)\lambda_2(\rho)$, we have the following equalities in qubit systems:
\begin{align}
V_\rho(A)V_\rho(B)
={}&
\frac14
\bm(
\Tr|\sqrt\rho[A,B]\sqrt\rho|
\bm)^2
+
\mathrm{Cov}_\rho(A,B)^2
\label{eq:QubitExactEqualityNagaokaHayashi}
\\
={}&
\frac14|\langle [A,B]\rangle_\rho|^2
+
\mathrm{Cov}_\rho(A,B)^2
+
\frac{1-P(\rho)}{2}\langle|[A,B]|^2\rangle_\rho
\label{eq:QubitExactEqualityKimuraMayumiOhno}
\\
={}&
\frac14|\langle [A,B]\rangle_\rho|^2
+
\mathrm{Cov}_\rho(A,B)^2
+
\frac{1-P(\rho)}{2^{1+2/p}}\|[A,B]\|^2_p
,
\quad
\forall\,p\in[1,\infty]
.
\label{eq:QubitExactEquality}
\end{align}

To see the separation of state dependence from
noncommutativity more clearly, we use the Bloch representation.
We write the observables and the state as $A=\frac12(\Tr A)\I+\bm a\cdot \bm\sigma$, $B=\frac12(\Tr B)\I+\bm b\cdot \bm\sigma$, and $\rho=\frac12(\I+\bm r\cdot\bm\sigma)$ 
with real vectors 
$\bm a,\bm b,\bm r\in\mathbb R^3$ and $\bm\sigma=(\sigma_x,\sigma_y,\sigma_z)$ denotes the vector of Pauli matrices, which satisfy $\Tr\sigma_i=0$ and $\sigma_i\sigma_j=\delta_{ij}\I+\mathrm i\sum_{k}\varepsilon_{ijk}\sigma_k$. 
Here, $\delta_{ij}$ is the Kronecker delta and $\varepsilon_{ijk}$ the Levi-Civita symbol.

Using the Bloch representations above, the additional term in our relation can be written as

\begin{equation}
\frac{\lambda_1(\rho)\lambda_2(\rho)}{4^{1/p}}\|[A,B]\|_p^2
=
\underbrace{(1-|\bm r|^2)}_{\text{state mixedness}}\times\underbrace{|\bm a\times \bm b|^2}_{\text{noncommutativity}}
\end{equation}
This expression explicitly separates the mixedness of the state, characterized by $(1-|\bm r|^2)$, from the noncommutativity of the observables, given by $|\bm a\times\bm b|^2$.
For qubit system, all three relations are equivalent. However, our Schatten norm term clearly shows the separate contributions of the state and the noncommutativity of the observables.


\subsection{Comparison in a qutrit system}

As discussed above, generalized uncertainty relations~(\ref{eq:NagaokaHayashi}) and (\ref{eq:KimuraMayumiOhno}), together with our result~(\ref{eq:MainResult}), all hold with equality in qubit systems. 

In higher-dimensional systems, however, these bounds do not generally coincide with the variance product, and their strength can depend on the choice of observables.
To compare these bounds, we consider an example in a qutrit system.
Since $p=1$ gives the strongest lower bound within our family of Schatten-$p$ norm relations, we focus on this case.
 
In the basis $\{\ket{1},\ket{2},\ket{3}\}$, consider the state $\rho$ and the observable $A_0$ given by
 \begin{equation}
     \rho=
     \begin{pmatrix}
     1/6 & 0 & 0
     \\
     0 & 1/3 & 0
     \\
     0 & 0 & 1/2
     \end{pmatrix}
     ,
     \quad
     A_0=
     \begin{pmatrix}
     1/3 & 0 & 0
     \\
     0 & -1/6 & 0
     \\
     0 & 0 & 0
     \end{pmatrix}
     .
\label{eq:QutritStateAndObservableA}
 \end{equation}
We consider two families of observables $B_s^{(1)}$ and $B_s^{(2)}$, depending on a parameter $s\in[0,1]$:
\begin{equation}
    B_s^{(1)}
    =
    \begin{pmatrix}
        0 & 1 & s
        \\
        1 & 0 & s
        \\
        s & s & 0
    \end{pmatrix}
    ,
    \quad
    B_s^{(2)}
    =
    \begin{pmatrix}
        0 & -\mathrm i & -\mathrm is
        \\
        \mathrm i & 0 & -\mathrm is
        \\
        \mathrm is & \mathrm is & 0
    \end{pmatrix}
    ,
    \quad
    s\in[0,1]
    .
    \label{eq:QutritObservableB}
\end{equation}

At $s=0$, the pair $(A_0,B_0^{(1)})$ is a special case of the construction given in Sec.~\ref{section:ProofMainTheorem} to attain equality in Eq.~\eqref{eq:MainResult}. 
The pair $(A_0,B_0^{(2)})$ is obtained by replacing $\ket{2}$ with $\mathrm{i}\ket{2}$ in the same construction. 
Thus, both pairs attain equality in Eq.~\eqref{eq:MainResult}. 
They also attain equality in Eqs.~\eqref{eq:KimuraMayumiOhno} and \eqref{eq:NagaokaHayashi}.

    The two choices of $B$ give the same variance: $V_\rho(B_s^{(1)})=V_\rho(B_s^{(2)})=1/2+3s^2/2$. Since $V_{\rho}(A_0)= 1/36$, the variance product is also the same for both choices: $V_{\rho}(A_0)V_{\rho}(B_s^{(i)})=1/72+s^2/24$.

\begin{figure}[t]
 \centering
        \includegraphics[width=\textwidth]{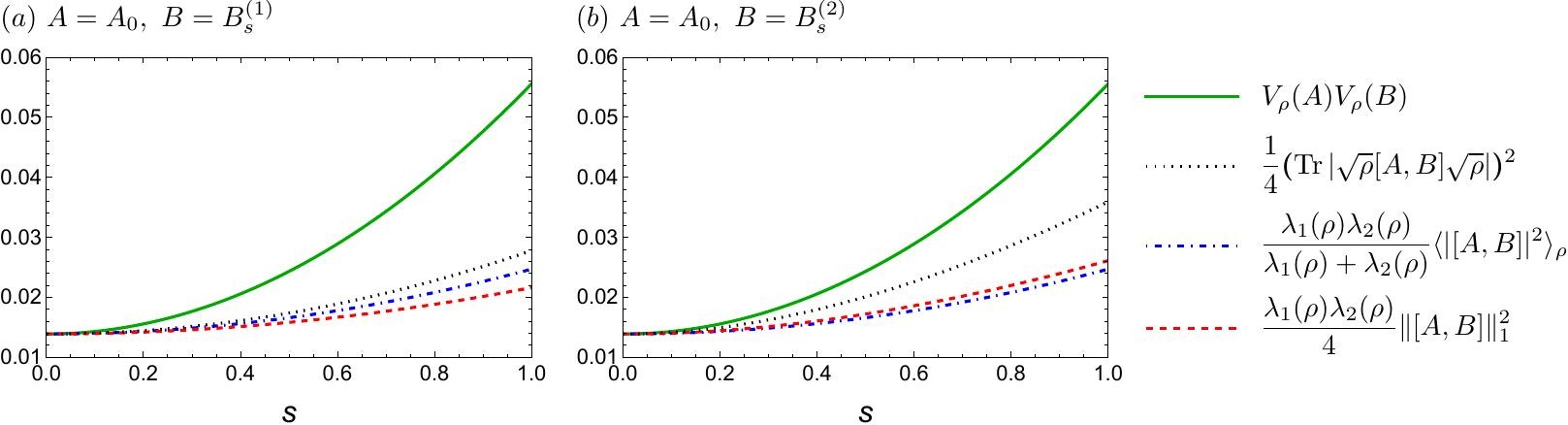}
 \caption{
 Comparison of the variance product and the three lower bounds in a qutrit system. 
 The state $\rho$ and the observable $A_0$ are given by Eq.~\eqref{eq:QutritStateAndObservableA}.
 The observable $B$ is chosen as $B_s^{(1)}$ in (a) and $B_s^{(2)}$ in (b), where $s\in[0,1]$, see Eq.~\eqref{eq:QutritObservableB}.
 The green solid, black dotted, blue dash-dotted, and red dashed lines represent the variance product $V_{\rho}(A_0)V_{\rho}(B_s^{(i)})$, the Nagaoka--Hayashi bound~\eqref{eq:NagaokaHayashi}, the KMOLC bound~\eqref{eq:KimuraMayumiOhno}, and our bound~\eqref{eq:MainResult} at $p=1$, respectively.
 For both choices of $B$, we have $\langle[A_0,B_s^{(i)}]\rangle_{\rho}=\Cov_{\rho}(A_0,B_s^{(i)})=0$.
 Thus, the Robertson--Schr\"{o}dinger relation gives a trivial lower bound.
 All three generalized uncertainty relations are saturated at $s=0$.
 }
\label{fig:ComparingURs}
\end{figure}

    The KMOLC bound in Eq.~\eqref{eq:KimuraMayumiOhno} also gives the same value for both choices. To see this, note that

\begin{equation}
    \bm|[A_0,B_s^{(1)}]\bm|^2
    =
    \begin{pmatrix}
        \frac14+\frac19s^2 & -\frac{1}{18}s^2 & \frac{1}{12}s
        \\
        -\frac{1}{18}s^2 & \frac14+\frac{1}{36}s^2 & \frac16s
        \\
        \frac{1}{12}s & \frac16s & \frac{5}{36}s^2
    \end{pmatrix}
    ,
    \quad
    \bm|[A_0,B_s^{(2)}]\bm|^2
    =
    \begin{pmatrix}
        \frac14+\frac19s^2 & -\frac{1}{18}s^2 & {\color{black}-\frac{1}{12}s}
        \\
        -\frac{1}{18}s^2 & \frac14+\frac{1}{36}s^2 & \frac16s
        \\
        {\color{black}-\frac{1}{12}s} & \frac16s & \frac{5}{36}s^2
    \end{pmatrix}
    \label{eq:NumericalExampleAbsoluteSquaredCommutators}
\end{equation}
Since $\rho$ is diagonal, only the diagonal entries of the matrices in Eq.~\eqref{eq:NumericalExampleAbsoluteSquaredCommutators} contribute to the expectation values. 
These entries are the same for both choices of $B$, we obtain
$\langle|[A_0,B_s^{(1)}]|^2\rangle_\rho=\langle|[A_0,B_s^{(2)}]|^2\rangle_\rho=1/8+7s^2/72$.
On the other hand, our optimal bound depends on the trace norm of commutator: $\|[A_0,B_s^{(i)}]\|_1= \Tr|[A_0,B_s^{(i)}]|$.
The two matrices in Eq.~\eqref{eq:NumericalExampleAbsoluteSquaredCommutators} differ in the signs of their $(1,3)$ and $(3,1)$ off-diagonal entries. 
A direct calculation shows that this difference leads to different trace norms for the two choices of $B$ when $s>0$, as shown in Fig.~\ref{fig:ComparingURs}.

These results show that the strength of our bound and the KMOLC bound depends on the choice of $B$. For $s>0$, the KMOLC bound is larger than our bound when $B=B_s^{(1)}$, as shown in Fig.~\ref{fig:ComparingURs}(a). 
In contrast, our bound is larger than the KMOLC bound when $B=B_s^{(2)}$, as shown in Fig.~\ref{fig:ComparingURs}(b). 
The Nagaoka--Hayashi bound gives the strongest lower bound in both cases. 
A general analysis of the relations among these bounds will be presented in a separate paper currently in preparation~\cite{KimuraEtAlInPreparation}.

\section{Conclusions}
\label{section:Conclusions}

   In this work, we have established a new family of uncertainty relations that separate state dependence from the noncommutativity of observables. 
   Our aim was to separate state dependence from noncommutativity rather than to remove it.
   To this end, we considered an optimization problem for each fixed state.
   
   Proposition~\ref{prop:MainTheorem} gives a solution to this problem for every $p\in[1,\infty]$: the optimal lower bound is given by the product of a coefficient depending only on the state and the squared Schatten-$p$ norm of the commutator. 
   In particular, the optimal coefficient is determined solely by the two smallest eigenvalues of the state. 
   Furthermore, we have shown that the same Schatten-$p$ term can be added to the Robertson--Schr\"odinger lower bound without reducing its optimal coefficient (Theorem~\ref{theorem:MainTheorem}). 
   This additional term is strictly positive for any full-rank state and any noncommuting pair of observables. 
   Our relation thus quantifies the noncommutative trade-off that is not captured by the conventional Robertson--Schr\"odinger relation.
   In qubit systems, our strengthened relation becomes an exact identity for arbitrary states and observables. 
   Although this property is also shared by the Nagaoka--Hayashi relation~\eqref{eq:NagaokaHayashi} and the KMOLC relation~\eqref{eq:KimuraMayumiOhno}, our formulation explicitly separates the mixedness of the state from the noncommutativity of the observables in the form $(1-|\bm r|^2)|\bm a\times\bm b|^2$. 
   This clarifies how the mixedness of the state and the noncommutativity of the observables contribute to the variance product.
   We have also shown that the trace-norm case $p=1$ gives the strongest lower bound within our families of Schatten-$p$ uncertainty relations. 
   However, our qutrit examples show that neither our trace-norm bound nor the KMOLC bound is always stronger than the other. 
   For future work, it would be interesting to clarify the differences between these generalized uncertainty relations, particularly in terms of their tightness.


\section*{Acknowledgments}
We are grateful to Izumi Tsutsui and Jaeha Lee for valuable discussions and comments.
K.K. and G.K. were supported by JSPS KAKENHI Grant Nos. JP24K22864 and JP24K06873, respectively.
The authors used ChatGPT (OpenAI; GPT-5.5 Thinking, GPT-5.6 Sol, GPT-6 Astra, and GPT-6.2) to assist with language editing, manuscript organization, and the development and verification of mathematical arguments. 
The authors reviewed and verified all AI-assisted outputs and take full responsibility for the content of the manuscript.

\section*{Data availability}
No experimental or external datasets were generated or analysed during the current study. All analytical results are established by rigorous mathematical proofs given in the main text and Appendices, together with the figure generated with numerical simulation.


\appendix

\section{Proof of Lemma~\ref{theorem:CalculatingWeightedNorm}}
\label{appendix:CalculatingWeightedNorm}

We begin by recalling Cauchy's interlacing theorem in an operator formulation suited to our setting (see, e.g., Ref.~\cite{Bhatia1997}).
\begin{proposition}[\textbf{Cauchy's interlacing inequality}]
\label{theorem:CauchyInterlacing}
Let $A\in\mathcal L_h(\mathcal H)$ and let $\{|\phi_1\rangle,|\phi_2\rangle,\ldots,|\phi_k\rangle\}$ be an orthonormal set in $\mathcal H$, with $1\leq k\leq d$. 
The compression of $A$ to the subspace $W=\operatorname{span}\{|\phi_1\rangle,\ldots,|\phi_k\rangle\}$ is the operator $A_k\in\mathcal L_h(W)$ defined by
\begin{equation}
A_k
=
\sum_{i=1}^k
\sum_{j=1}^k
\langle \phi_i|A|\phi_j\rangle
|\phi_i\rangle\langle\phi_j|
\end{equation}
Let $\lambda_1(A)\leq\cdots\leq\lambda_d(A)$ and $\lambda_1(A_k)\leq\cdots\leq\lambda_k(A_k)$ denote the eigenvalues of $A$ and $A_k$, respectively, counted with multiplicity. 
Then
\begin{equation}
\lambda_l(A)
\leq
\lambda_{l}(A_k)
\leq
\lambda_{d-k+l}(A)
\quad
(l=1,2,\ldots,k)
.
\label{eq:CauchyInterlacing}
\end{equation}
\end{proposition}

The following lemma is obtained by applying Cauchy's interlacing theorem to a two-dimensional compression and estimating its determinant.
\begin{lemma}
\label{theorem:CauchyInterlacingCorollary}
Let $\omega>0$ be a positive-definite operator and let $\{|\phi\rangle,|\psi\rangle\}$ be an orthonormal set. Then, for every $z\in\mathbb C$, we have
\begin{equation}
\langle \psi+z\phi|\omega|\psi+z\phi\rangle
\geq
\frac{\lambda_1(\omega)\lambda_2(\omega)}
{\langle \phi|\omega|\phi\rangle}.
\end{equation}
\end{lemma}

\begin{proof}
Let $\omega_2$ be the compression of $\omega$ to
$W=\operatorname{span}\{|\psi\rangle,|\phi\rangle\}$ with the matrix representation
\begin{equation}
\omega_2
=
\langle \psi|\omega|\psi\rangle|\psi\rangle\langle\psi|
+
\langle \psi|\omega|\phi\rangle|\psi\rangle\langle\phi|
+
\langle \phi|\omega|\psi\rangle|\phi\rangle\langle\psi|
+
\langle \phi|\omega|\phi\rangle|\phi\rangle\langle\phi|.
\end{equation}
By the interlacing inequality~(\ref{eq:CauchyInterlacing}), $0<\lambda_1(\omega)\leq\lambda_1(\omega_2)$ and $0<\lambda_2(\omega)\leq\lambda_2(\omega_2)$, which implies
\begin{equation}
\lambda_1(\omega)\lambda_2(\omega)
\leq
\lambda_1(\omega_2)\lambda_2(\omega_2)
=
\det \omega_2
=
\langle\psi|\omega|\psi\rangle\langle\phi|\omega|\phi\rangle
-|\langle\psi|\omega|\phi\rangle|^2.
\end{equation}
For any $z\in\mathbb C$, completing the square gives
\begin{align}
\langle \psi+z\phi|\omega|\psi+z\phi\rangle
={}&
\langle\psi|\omega|\psi\rangle
+2\Re\bm(z\langle\psi|\omega|\phi\rangle\bm)
+|z|^2\langle \phi|\omega|\phi\rangle
\nonumber\\
={}&
\langle\psi|\omega|\psi\rangle
+
\langle \phi|\omega|\phi\rangle
\biggl|
z+\frac{\langle\psi|\omega|\phi\rangle^\ast}{\langle\phi|\omega|\phi\rangle}
\biggr|^2
-
\frac{|\langle\psi|\omega|\phi\rangle|^2}{\langle\phi|\omega|\phi\rangle}
\nonumber\\
\geq{}&
\frac{1}{\langle\phi|\omega|\phi\rangle}
\bm(
\langle\psi|\omega|\psi\rangle
\langle\phi|\omega|\phi\rangle
-
|\langle\psi|\omega|\phi\rangle|^2
\bm)
\geq
\frac{\lambda_1(\omega)\lambda_2(\omega)}{\langle\phi|\omega|\phi\rangle}.
\end{align}
This proves the claim.
\end{proof}

We are now ready to prove Lemma~\ref{theorem:CalculatingWeightedNorm}.  

\begin{proof}[Proof of Lemma~\ref{theorem:CalculatingWeightedNorm}]
Let $\omega>0$ be a positive-definite operator, with  its two smallest eigenvalues $\lambda_1(\omega)$ and $\lambda_2(\omega)$. 
Let $X\in\mathcal L(\mathcal H)$ and let $\{|\phi_i\rangle\}_{i=1}^d$ be an orthonormal basis of $\mathcal H$.
Write $X_{ij}=\langle\phi_i|X|\phi_j\rangle$ and $\omega_{ii}=\langle\phi_i|\omega|\phi_i\rangle>0$.

For each $i$, define
\begin{equation}
C_i=\left(\sum_{j\neq i}|X_{ij}|^2\right)^{1/2}.
\end{equation}
If $C_i>0$, the vector
\begin{equation}
|\psi_i\rangle
=
\frac{1}{C_i}
\sum_{j\neq i}X_{ij}^\ast|\phi_j\rangle
\end{equation}
is a unit vector orthogonal to $|\phi_i\rangle$.
Setting $z_i=X_{ii}^\ast/C_i$, we have
\begin{equation}
X^\dagger|\phi_i\rangle
=
X_{ii}^\ast|\phi_i\rangle
+
\sum_{j\neq i}X_{ij}^\ast|\phi_j\rangle
=
C_i\bigl(z_i|\phi_i\rangle + |\psi_i\rangle\bigr),
\end{equation}
and thus
\begin{equation}
\langle\phi_i|X\omega X^\dagger|\phi_i\rangle
=C_i^2\langle\psi_i+z_i\phi_i|\omega|\psi_i+z_i\phi_i\rangle.
\end{equation}
Lemma~\ref{theorem:CauchyInterlacingCorollary} therefore gives
\begin{equation}
\langle\phi_i|X\omega X^\dagger|\phi_i\rangle
\geq
C_i^2
\frac{\lambda_1(\omega)\lambda_2(\omega)}{\omega_{ii}}.
\end{equation}
If $C_i=0$, then $X^\dagger|\phi_i\rangle=X_{ii}^\ast|\phi_i\rangle$, so
\begin{equation}
\langle\phi_i|X\omega X^\dagger|\phi_i\rangle
=
|X_{ii}|^2\omega_{ii}\geq0 =
C_i^2\frac{\lambda_1(\omega)\lambda_2(\omega)}{\omega_{ii}}.
\end{equation}
Thus, the same lower bound holds for every $i$.
Using the cyclicity of the trace and summing over $i$, we obtain
\begin{align}
\langle|X|^2\rangle_\omega
&=
\Tr(\omega X^\dagger X)
=
\sum_{i=1}^d
\langle\phi_i|X\omega X^\dagger|\phi_i\rangle
\nonumber\\
&\geq
\lambda_1(\omega)\lambda_2(\omega)
\sum_{i=1}^d\frac{C_i^2}{\omega_{ii}}
\nonumber\\
&=
\lambda_1(\omega)\lambda_2(\omega)
\sum_{i\neq j}\frac{|X_{ij}|^2}{\omega_{ii}},
\end{align}
which proves the claim.
\end{proof}


\section{Proof of inequality~(\ref{eq:YNormEvaluation})}
\label{appendix:ProofOfYNormEvaluation}

We prove the squared inequality of~(\ref{eq:YNormEvaluation}) just for convenience. 
Recall that $Y$ is Hermitian, $\omega>0$ is positive definite, and $U$ is unitary.
Let $\{|\phi_i\rangle\}_{i=1}^d$ be an orthonormal eigenbasis of $Y$, and write
$\omega_{ii}=\langle\phi_i|\omega|\phi_i\rangle>0$ and
$U_{ij}=\langle\phi_i|U|\phi_j\rangle$.
By the Cauchy--Schwarz inequality,
\begin{equation}
|\lambda_i(Y)-\lambda_j(Y)|^2
\leq
\left(\frac{1}{\omega_{ii}}+\frac{1}{\omega_{jj}}\right)
\left(\omega_{ii}\lambda_i(Y)^2+\omega_{jj}\lambda_j(Y)^2\right).
\end{equation}
Consequently,
\begin{align}
&\sum_{\lambda_i(Y)\neq\lambda_j(Y)}
\left(\frac{1}{\omega_{ii}}+\frac{1}{\omega_{jj}}\right)^{-1}
|U_{ij}|^2|\lambda_i(Y)-\lambda_j(Y)|^2
\nonumber\\
&\quad\leq
\sum_{i,j=1}^d
|U_{ij}|^2
\left(\omega_{ii}\lambda_i(Y)^2+\omega_{jj}\lambda_j(Y)^2\right)
\nonumber\\
&\quad=
\sum_{i=1}^d
\omega_{ii}\lambda_i(Y)^2
\sum_{j=1}^d|U_{ij}|^2
+
\sum_{j=1}^d
\omega_{jj}\lambda_j(Y)^2
\sum_{i=1}^d|U_{ij}|^2
\nonumber\\
&\quad=
2\sum_{i=1}^d\omega_{ii}\lambda_i(Y)^2
=
2\Tr(\omega Y^2)
=
2\langle|Y|^2\rangle_\omega.
\end{align}
Here, we extended the sum to all pairs $(i,j)$ by adding nonnegative terms and used
\begin{equation}
\sum_{j=1}^d|U_{ij}|^2
=
\langle\phi_i|UU^\dagger|\phi_i\rangle
=
1,
\qquad
\sum_{i=1}^d|U_{ij}|^2
=
\langle\phi_j|U^\dagger U|\phi_j\rangle
=
1.
\end{equation}


\section{Further observation of Lemma~\ref{theorem:TraceNormDominance} and its extension}
\label{appendix:ProofOfTraceNormDominanceGeneralized}

In this appendix, we examine inequality~\eqref{eq:TraceNormDominance} for traceless Hermitian operators in qubit and qutrit systems.
We also establish a related inequality for arbitrary linear operators, clarifying the role of the traceless Hermitian structure in Lemma~\ref{theorem:TraceNormDominance}.

For qubit systems, the following identity explains why the additional term in our uncertainty relations is independent of $p$.

\begin{lemma}
\label{lemma:QubitTraceNormIdentity}
Let $H$ be a traceless Hermitian operator on a two-dimensional Hilbert space.
Then, for every $p\in[1,\infty]$,
\begin{equation}
2^{1-1/p}\|H\|_p=\|H\|_1.
\end{equation}
\end{lemma}

\begin{proof}
Since $H \in {\mathcal L}({\mathbb C}^2)$ is traceless and Hermitian, its eigenvalues are $-a$ and $a$ for some $a\geq0$.
Thus, $|H|=a\I$, so
\begin{equation}
\|H\|_p= \bigl(|a|^p + |-a|^p\bigr)^{1/p} = 
2^{1/p}a
\quad (1\leq p<\infty),
\qquad
\|H\|_\infty=a,
\qquad
\|H\|_1=2a.
\end{equation}
The claimed identity follows\footnote{Alternatively, this identity also follows directly from the proof of Lemma~\ref{theorem:TraceNormDominance}, since each of the positive and negative parts of $H$ has at most one nonzero eigenvalue.}.
\end{proof}

For qutrit systems, the spectral-norm case ($p=\infty$) always yields the same lower bound as the trace-norm case ($p=1$).

\begin{lemma}
Let $H$ be a traceless Hermitian operator on a three-dimensional Hilbert space.
Then
\begin{equation}
2\|H\|_\infty=\|H\|_1.
\end{equation}
\end{lemma}
\begin{proof}
The claim is trivial if $H=0$.
Otherwise, replacing $H$ by $-H$ if necessary, we may assume that
\begin{equation}
\lambda_1(H)<0\leq\lambda_2(H)\leq\lambda_3(H).
\end{equation}
Since $\Tr H=0$, we have
\begin{equation}
|\lambda_1(H)|=\lambda_2(H)+\lambda_3(H).
\end{equation}
Thus, $|\lambda_1(H)|\geq\lambda_3(H)$, and hence
\begin{equation}
\|H\|_\infty= \max\{|\lambda_1(H)|,|\lambda_2(H)|,|\lambda_3(H)|\}
=
|\lambda_1(H)|
=
\frac12\left(
|\lambda_1(H)|+\lambda_2(H)+\lambda_3(H)
\right)
=
\frac12\|H\|_1.
\end{equation}
This proves the claim.
\end{proof}

In arbitrary finite dimensions, the trace norm need not maximize the quantity $2^{1-1/p}\|X\|_p$ for a general linear operator $X$.
The following proposition shows that the maximum over $p\in[1,\infty]$ is always attained at $p=1$ or $p=\infty$.

\begin{proposition}

Let $X$ be an arbitrary linear operator, then
\begin{equation}
2^{1-1/p}
\|X\|_p
\leq
\max
\{\|X\|_1,2\|X\|_\infty\}
\quad
(1\leq p\leq \infty)
.
\label{eq:GeneralizedTraceNormDominance}
\end{equation}

\end{proposition}

\begin{proof}

This result essentially comes from the convexity of the following function:
\begin{equation}
f_X(q)
=
(1-q)\log 2
+
\log\|X\|_{1/q}
,
\quad
q\in[0,1]
.
\end{equation}
Indeed, the interpolation inequality between Schatten norms~\cite{Bhatia1997}
\begin{equation}
\|X\|_{p(\theta)}
\leq
\|X\|_{p_0}^{1-\theta}\|X\|_{p_1}^\theta
,
\quad
\frac{1}{p(\theta)}
=
\frac{1-\theta}{p_0}+\frac{\theta}{p_1},
\quad
p_0,p_1\in[1,\infty),
\quad
\theta\in[0,1]
,
\end{equation}
ensures the convexity of $\log \|X\|_{1/q}$ on $q\in(0,1]$ by substituting 
$q_0=1/p_0$, $q_1=1/p_1$, that is,
\begin{equation}
\log \|X\|_{1/\bm((1-\theta)q_0+\theta q_1\bm)}
\leq
(1-\theta)\log \|X\|_{1/q_0}
+
\theta\log
\|X\|_{1/q_1}
,
\quad
q_0,q_1\in(0,1]
.
\end{equation}
Moreover, we can extend this statement from $q_0,q_1\in(0,1]$ to $q_0,q_1\in[0,1]$ since 
the one-sided limit $\lim_{q\to0+}\|X\|_{1/q}=\|X\|_\infty$ exists. 
Therefore $f_X(q)$ is the convex function satisfying
\begin{align}
\log\bm(
2^{1-q}\|X\|_{1/q}
\bm)
=
f_X(q)
\leq{}&
(1-q)f_X(0)
+
qf_X(1)
\nonumber\\
\leq{}&
\max\{f_X(0),f_X(1)\}
=
\max\{\log \bm(2\|X\|_\infty\bm),\log \|X\|_1\}
,
\end{align}
which implies the desired result~(\ref{eq:GeneralizedTraceNormDominance}) by resubstituting $p=1/q\in[1,\infty]$.

\end{proof}

\begin{remark}
For qubit systems we have $\|X\|_1=\sigma_1(X)+\sigma_2(X)\leq 2\|X\|_\infty$ and therefore $2^{-1/p}\|X\|_p\leq \|X\|_\infty$ for all $p\in[1,\infty]$, i.e., the dominance of spectral norm instead of the trace norm. 
In higher-dimensional systems, the optimality depends on the cases.
Indeed, the trace norm becomes optimal to measure the identity operator $\I=\mathrm{diag}(1,1,1)$ on qutrit systems as
$\|\I\|_1=3>2=2\|\I\|_\infty$.
\end{remark}


\bibliographystyle{apsrev4-2}
\bibliography{StateIndependentURs}

\end{document}